\documentstyle[aps]{revtex}
\makeatletter

          \@addtoreset{equation}{section}
       \makeatother
\begin{document}
\draft
\title{%
BRST invariant Lagrangian of spontaneously broken gauge theories \\
in noncommutative geometry
}
\author{%
  Yoshitaka {\sc Okumura}\thanks{
e-mail address: okum@isc.chubu.ac.jp}}
\address{Department of Natural Science, 
Chubu University, Kasugai, 487, Japan
}
\maketitle
\vskip 0.5cm
\begin{abstract}
The quantization of spontaneously broken gauge theories in  
noncommutative geometry(NCG) has been  sought for some time,
because quantization is crucial for making the NCG approach a
reliable and physically acceptable theory.
Lee, Hwang and Ne'eman  recently succeeded  in realizing the BRST 
quantization of  gauge theories
 in NCG in the matrix derivative approach proposed by Coquereaux et al.
The present author has  proposed a characteristic formulation to 
reconstruct a gauge theory in NCG on the discrete space $M_4\times Z_{_N}$.
Since this formulation is a generalization 
of the differential geometry on the ordinary
manifold to that on the discrete manifold, it is more familiar than 
other approaches. In this paper, we show that within 
our formulation we can obtain the BRST invariant 
Lagrangian in the same way as Lee, Hwang and Ne'eman and apply it to 
the SU(2)$\times$U(1) gauge theory.  
\vskip 0.3cm
\noindent
PACS number(s):11.15.Ex,02.40.Hw,12.15-y
\end{abstract}

\section{\bf Introduction}
\label{intro}
The effort made to understand the Higgs mechanism has  produced 
 many attempts
including the Kaluza-Klein model\cite{MM}, \cite{Man}, the technicolor
 model,  
top quark condensation\cite{YK}, 
and the approach based on noncommutative geometry(NCG) on the discrete
 space. 
Among these, the NCG approach is most promising because 
the reason for the existence 
of the Higgs field becomes apparent, 
and no extra physical modes are necessary. 
Since Connes\cite{Con} proposed the first idea concerning NCG, 
many works\cite{MM}$\sim$\cite{Soga} have 
appeared, realizing the unified picture of gauge and Higgs fields as 
the generalized connection on the discrete space $M_4\times Z_2.$\par
We have also presented a characteristic
 formulation \cite{MO1}$\sim$\cite{O10} which  
is the generalization of the usual differential geometry on an 
ordinary manifold to the discrete space $M_4\times Z_{_N}$.
In noncommutative geometry on $M_4\times Z_2$,
the extra differential one-form $\chi$ 
is introduced in addition to
the usual one-form $dx^\mu$, and therefore our formulation is very similar
to ordinary differential geometry. In contrast, the  original 
formulation by Connes is
very difficult to understand. The one-form basis $\chi$ was originally 
introduced by Sitarz\cite{Sita}. 
However, his scheme is somewhat difficult to 
apply to the reconstruction of the model in particle physics
such as the Standard Model
and Grand Unified Model(GUT). We improved Sitarz's scheme by introducing 
a symmetry breaking matrix to enable us to reconstruct
 gauge theories  such as the Standard Model\cite{OFS}, the left-right 
symmetric gauge theory\cite{LR}, 
SU(5) GUT\cite{MO2}, and SO(10) GUT\cite{O10}. \par
Among the many attempts to reconstruct the Standard Model 
in NCG, the gauge invariant
Lagrangian for bosonic and fermionic sectors has been
 obtained. 
However, the way to construct the gauge fixing term 
and ghost term in the NCG scheme to this point has not been
evident, though these terms are very important 
to ensure the quantization of the gauge theory.  Until now, these terms 
have been added by hand, without justification by 
any reasonable method in NCG. 
Lee, Hwang and Ne'eman \cite{LHN} succeeded 
very recently in obtaining the BRST quantization of a gauge 
theory in NCG in the matrix derivative approach due 
to Coquereaux and others \cite{CHPS}.
They obtained the BRS/anti-BRS transformation rules of the theory by
applying the horizontality condition in the super field formulation
and constructed BRST invariant Lagrangian including the gauge fixing and
ghost terms. This construction  yielded two important features,
 that the t'Hooft gauge is obtained and  
 that the odd part in matrix formulation produces a global symmetry.
\par
In this article, we apply  similar method to our formulation
in order to obtain the BRST invariant Lagrangian 
of the spontaneously broken gauge
theory in NCG. We find results similar to those of 
Lee, Hwang and Ne'eman \cite{LHN}. 
The fermion sector is not treated because it is not related to 
the main theme of this article.
\par
This article consists of four sections. The next section presents 
the general settings of our formulation by introducing the Grassmann
 number
$\theta$ and $\bar {\theta}$ in addition
 to $x_\mu$ in $M_4$ and $y$ in the 
discrete space $Z_2$. Ghost fields are given as  members of
generalized gauge field in the same way as gauge and Higgs fields.
The horizontality condition is applied to obtain 
the BRS/anti-BRS transformation(BRST/anti-BRST)of the respected fields, 
and the method to obtain the gauge fixing and ghost terms is presented.
The third section is an application to the SU(2)$\times$U(1) gauge model,
where results very similar to those of  \cite{LHN} are obtained.
The last section is devoted to concluding remarks. 
\section{  General settings}
\label{sec.2}
In the previous formulation \cite{MO1}, \cite{OFS}, 
we adopted the discrete space $M_4\times Z_2$
to reconstruct the Standard Model.
According to the super
field formulation of Bonora and Tonin\cite{BT},
 the Grassmann numbers $\theta$
and $\bar{\theta}$ are added to $x_\mu$ and $y$ in $M_4\times Z_2$
to produce the ghost and 
anti-ghost fields. However, we do not identify
 the super-space of arguments 
$\theta$ and $\bar{\theta}$ because this
 is not necessary to obtain the final
results.\par
Let us start with the equation of the generalized gauge field 
${\cal A}(x,y,\theta,\bar{\theta})$
written in one-form on the discrete space $M_4\times Z_2$:
\begin{equation}
      {\cal A}(x,y,\theta,\bar{\theta})
      =\sum_{i}a^\dagger_{i}(x,y,\theta,\bar{\theta}){\bf d}
      a_i(x,y,\theta,\bar{\theta}). \label{2.1}
\end{equation}
Here, the $a_i(x,y,\theta,\bar{\theta})$ are
 the square-matrix-valued functions.
The subscript $i$ is  a variable corresponding to the extra
internal space which we cannot now identify. 
At this time, we simply regard $a_i(x,y,\theta,\bar{\theta})$  
as a more fundamental
field from which to construct gauge and Higgs fields,
 because they have only
a mathematical meaning. The functions $a_i(x,y,\theta,\bar{\theta})$ 
do not appear in the final stage.
The operator ${\bf d}$ in Eq.(\ref{2.1}) is the generalized exterior 
derivative defined as follows:
\begin{eqnarray} 
  &&     {\bf d}=d + d_\chi+d_\theta+ d_{\bar{\theta}}, \label{2.2a1}\\  
  &&   da_i(x,y,\theta,\bar{\theta}) = 
   \partial_\mu a_i(x,y,\theta,\bar{\theta})dx^\mu, \label{2.2a2}\\
  && d_{\chi} a_i(x,y,\theta,\bar{\theta}) 
    =\partial_y a_i(x,y,\theta,\bar{\theta})\chi 
    \nonumber\\
    && \hskip 2.5cm=[-a_i(x,y,\theta,\bar{\theta})M(y) 
      + M(y)a_i(x,-y,\theta,\bar{\theta})]\chi, \label{2.2a3}\\
  && d_\theta a_i(x,y,\theta,\bar{\theta})
       =\partial_\theta a_i(x,y,\theta,\bar{\theta})d\theta
         \label{2.2a4}\\
  &&  d_{\bar{\theta}} a_i(x,y,\theta,\bar{\theta})
    =\partial_{\bar{\theta}}a_i(x,y,\theta,\bar{\theta})d\bar{\theta}, 
    \label{2.2a5}
\end{eqnarray}
where the derivative $\partial_y$ is defined in Eq.(\ref{2.2a3}),
$dx^\mu$ is 
ordinary one-form basis, taken to be dimensionless, in Minkowski space 
$M_4$, and $\chi$ 
is the one-form basis, also assumed to be dimensionless, 
in the discrete space $Z_2$.  
The operators $d_\theta$ and $d_{\bar{\theta}}$ are also one-form
bases in super-space.
We have introduced the $x$-independent matrix $M(y)$, 
whose hermitian conjugation is given by $M(y)^\dagger=M(-y)$. 
The matrix $M(y)$ turns out to determine the scale and pattern of 
the spontaneous breakdown of the gauge symmetry. 
In order to find the explicit forms of gauge, Higgs and ghost fields
according to Eqs. (\ref{2.1}) and (\ref{2.2a1})$\sim$(\ref{2.2a5}), 
we need the following important algebraic rule of noncommutative geometry:
\begin{equation}
  f(x,y,\theta,\bar{\theta})\chi=\chi f(x,-y,\theta,\bar{\theta}), 
  \label{2.4}
\end{equation}
where $f(x,y,\theta,\bar{\theta})$ is a field such as
$a_i(x,y,\theta,\bar{\theta})$, a gauge, Higgs, ghost 
or  fermion field defined on the discrete space.
It should be noticed that Eq.(\ref{2.4}) does not express 
the relation between
the matrix elements of $f(x,+,\theta,\bar{\theta})$ 
and $f(x,-,\theta,\bar{\theta})$, but rather insures the product between
the fields expressed in  differential form on the discrete space. 
Its justification can be easily seen in the calculation
 of the wedge product
${\cal A}(x,y,\theta,\bar{\theta})\wedge
 {\cal A}(x,y,\theta,\bar{\theta})$. 
Equation(\ref{2.4}) realizes the non-commutativity of our algebra 
in the geometry on the discrete space $M_4\times Z_2$. 
Inserting Eq.(\ref{2.2a1})$\sim$Eq.(\ref{2.2a5}) into Eq.(\ref{2.1})
and using Eq.(\ref{2.4}),
${\cal A}(x,y,\theta,\bar{\theta})$ is rewritten as
\begin{equation}
 {\cal A}(x,y,\theta,\bar{\theta})={A}_\mu(x,y,\theta,\bar{\theta})dx^\mu
 +{\mit\Phi}(x,y,\theta,\bar{\theta})\chi
 +{C}(x,y,\theta,\bar{\theta})d\theta
 +{{\bar C}}(x,y,\theta,\bar{\theta})d{\bar\theta}, \label{2.5}
\end{equation}
where
\begin{eqnarray}
&&    A_\mu(x,y,\theta,\bar{\theta}) = \sum_{i}a_{i}^\dagger(x,y,
  \theta,\bar{\theta})\partial_\mu a_{i}(x,y,\theta,\bar{\theta}),  
  \label{2.61}\\
&&     {\mit\Phi}(x,y,\theta,\bar{\theta}) = \sum_{i}a_{i}^\dagger(x,y,
  \theta,\bar{\theta})\,\left[-a_i(x,y,\theta,\bar{\theta})M(y) 
            + M(y)a_i(x,-y,\theta,\bar{\theta})\right], \label{2.62}\\
&&  {C}(x,y,\theta,\bar{\theta})
  =\sum_{i}a_{i}^\dagger(x,y,\theta,\bar{\theta}) 
      \partial_\theta a_{i}(x,y\theta,\bar{\theta})
  \label{2.63}\\
&&  {\bar C}(x,y,\theta,\bar{\theta})
  =\sum_{i}a_{i}^\dagger(x,y,\theta,\bar{\theta}) 
      \partial_{\bar\theta} a_{i}(x,y,\theta,\bar{\theta}).
  \label{2.64}
\end{eqnarray}
Here, $A_\mu(x,y,\theta,\bar{\theta})$,
 ${\mit\Phi}(x,y,\theta,\bar{\theta})$, 
$C(x,y,\theta,\bar{\theta})$ and 
${\bar C}(x,y,\theta,\bar{\theta})$ are identified with
the gauge field in the flavor symmetry, Higgs field,
ghost and anti-ghost fields, respectively. 
In order to identify  $A_\mu(x,y,\theta,\bar{\theta})$ 
with true gauge fields, the following conditions must be imposed:
\begin{eqnarray}
&&    \sum_{i}a_{i}^\dagger(x,y,\theta,\bar{\theta})
     a_{i}(x,y,\theta,\bar{\theta})= 1. 
  \label{2.7}
\end{eqnarray}
Equation(\ref{2.7}) reminds us of the effective gauge field of the
Berry phase\cite{Ber}, though the parameter space
 of $i$ and Minkowski space of
$x^\mu$ are reversed. This might be a key to identifying the fundamental
field $a_{i}(x,y,\theta,\bar{\theta})$. Equation (\ref{2.7})
 is very important
and we use it often below. For later convenience, we define the one-form
fields as
\begin{eqnarray}
 &&  {\hat A}(x,y,\theta,\bar{\theta})
 =A_\mu(x,y,\theta,\bar{\theta})dx^\mu,
                     \label{2.6a1} \\
&&  {\hat{\mit\Phi}}(x,y,\theta,\bar{\theta})
={\mit\Phi}(x,y,\theta,\bar{\theta})\chi,
                     \label{2.6a2} \\
&&      {\hat C}(x,y,\theta,\bar{\theta})
={C}(x,y,\theta,\bar{\theta})d\theta, \label{2.6a3}\\ 
&&      {\hat{\bar C}}(x,y,\theta,\bar{\theta}),
               ={\bar C}(x,y,\theta,\bar{\theta})d{\bar\theta}
               .\label{2.6a4} 
\end{eqnarray}
\par 
Before constructing the gauge covariant field strength, 
we address the gauge transformation 
of $a_i(x,y,\theta,\bar{\theta})$,  which is defined as 
\begin{eqnarray}
&&      a^{g}_{i}(x,y,\theta,\bar{\theta})
        = a_{i}(x,y,\theta,\bar{\theta})g(x,y), 
\label{2.8}
\end{eqnarray}
where
$g(x,y)$ is the gauge function with respect to the corresponding
flavor unitary group.
Then, we can find from Eqs.(\ref{2.1}) and (\ref{2.8})
the gauge transformation of ${\cal A}(x,y,\theta,\bar{\theta})$ to be
\begin{equation}
{\cal A}^g(x,y,\theta,\bar{\theta})=g^{-1}(x,y) 
{\cal A}(x,y,\theta,\bar{\theta})g(x,y)
 +g^{-1}(x,y){\bf d}g(x,y), \label{2.9}
\end{equation}
where, as in Eq.(\ref{2.2a1})$\sim$Eq.(\ref{2.2a3}),
\begin{eqnarray} 
       {\bf d}g(x,y)&&=(d+d_\chi) g(x,y)=\partial_\mu g(x,y)dx^\mu+
                             \partial_yg(x,y)\chi \nonumber\\
             && = \partial_\mu g(x,y)dx^\mu+[-g(x,y)M(y)
              + M(y)g(x,-y)]\chi.
          \label{2.10}
\end{eqnarray}
Using Eqs.(\ref{2.8}) and (\ref{2.9}), 
we can find the gauge transformations of gauge, Higgs,
 ghost and anti-ghost
fields as
\begin{eqnarray}
&&       A_\mu^g(x,y,\theta,\bar{\theta})=g^{-1}(x,y)
   A_\mu(x,y,\theta,\bar{\theta})g(x,y)+
                           g^{-1}(x,y)\partial_\mu g(x,y),  
                           \label{2.11}\\
&&  {\mit\Phi}^g(x,y,\theta,\bar{\theta})
=g^{-1}(x,y){\mit\Phi}(x,y,\theta,\bar{\theta})
g(x,-y)+  g^{-1}(x,y)\partial_y g(x,y), \label{2.12}\\
&&      C^g(x,y,\theta,\bar{\theta})=g^{-1}(x,y)
C(x,y,\theta,\bar{\theta})
   g(x,y),  \label{2.13}\\
&&      {\bar C}^g(x,y,\theta,\bar{\theta})=g^{-1}(x,y)
      {\bar C}(x,y,\theta,\bar{\theta})g(x,y).  \label{2.13a}
\end{eqnarray}
Equation(\ref{2.12}) is very similar to Eq.(\ref{2.11}), 
the gauge transformation of
the genuine gauge field $A_\mu(x,y,\theta,\bar{\theta})$.  
This strongly suggests that the Higgs field is a kind of gauge field
on the discrete space $M_4\times Z_2$. From Eq.(\ref{2.10}),  
Eq.(\ref{2.12})  is rewritten as
\begin{equation}
       {\mit\Phi}^g(x,y,\theta,\bar{\theta})+M(y)=g^{-1}(x,y)
       ({\mit\Phi}(x,y,\theta,\bar{\theta})+M(y))g(x,-y)
                              \label{2.14}\\
\end{equation}
which makes obvious the fact that $H(x,y,\theta,\bar{\theta})$ defined as
\begin{equation}
H(x,y,\theta,\bar{\theta})={\mit\Phi}(x,y,\theta,\bar{\theta})
+M(y) \label{2.15}
\end{equation}
is the un-shifted Higgs field, whereas ${\mit\Phi}
(x,y,\theta,\bar{\theta})$ 
denotes the shifted Higgs field with  vanishing vacuum expectation value.
Equations (\ref{2.13}) and (\ref{2.13a})
 demonstrate that ghost and anti-ghost
fields are transformed as the field in adjoint representation.
\par
In addition to the algebraic rules
 in Eqs.(\ref{2.2a1})$\sim$(\ref{2.2a5}), 
we add one more important rule,
\begin{equation}
              d_\chi M(y)=M(y)M(-y)\chi,            \label{2.16}
\end{equation}
which together with Eq.(\ref{2.2a3}) yields  the nilpotency
of $\chi$ and, in turn, the nilpotency
 of the generalized exterior derivative 
$\bf d$ :
\begin{equation}
          {\bf d}^2 f(x,y,\theta,\bar{\theta})
          =(d^2+d_\chi^2+d^2_\theta+d^2_{\bar\theta})
          f(x,y,\theta,\bar{\theta})=0,  \label{2.17}
\end{equation}
with the natural conditions  
\begin{eqnarray}
&& dx^\mu\wedge\chi=-\chi\wedge dx^\mu, \hskip 1cm
   dx^\mu\wedge d\theta=-d\theta\wedge dx^\mu, \hskip 1cm
   dx^\mu\wedge d{\bar\theta}=-d{\bar\theta}\wedge dx^\mu, \nonumber\\
&& \chi\wedge d\theta=-d\theta\wedge \chi, \hskip 1cm  
   \chi\wedge d{\bar\theta}=-d{\bar\theta}\wedge \chi, \nonumber\\
&& d{\theta}\wedge d{\bar\theta}=d{\bar\theta}\wedge d\theta, \hskip 1cm
  \partial_\theta\partial_{\bar\theta}
  =-\partial_{\bar\theta}\partial_\theta.
      \label{2.16a}
\end{eqnarray}
For  proof of the nilpotency of $d_\chi$, see Ref.\cite{MO1}.
With these considerations we can construct the gauge covariant field
strength:
\begin{equation}
  {\cal F}(x,y,\theta,\bar{\theta})
  = {\bf d}{\cal A}(x,y,\theta,\bar{\theta})
  +{\cal A}(x,y,\theta,\bar{\theta})\wedge{\cal A}
  (x,y,\theta,\bar{\theta}).
\label{2.18}
\end{equation}
From Eqs.(\ref{2.9}) and (\ref{2.17}) we  easily find the gauge 
transformation of ${\cal F}(x,y,\theta,\bar{\theta})$ as
\begin{equation}
         {\cal F}^g(x,y,\theta,\bar{\theta})
         =g^{-1}(x,y){\cal F}(x,y,\theta,\bar{\theta})g(x,y).
           \label{2.20}
\end{equation}
\par
Here, following Bonora and Tonin\cite{BT}, we impose the horizontality
condition on ${\cal F}(x,y,\theta,\bar{\theta})$:
\begin{equation}
    {\cal F}(x,y,\theta,\bar{\theta})|_{\theta={\bar\theta}=0}
    =F(x,y), \label{2.20a}
\end{equation}
where $F(x,y)$ is the generalized field strength not accompanying
the one-form bases $d\theta$ and $d{\bar\theta}$.
Equation(\ref{2.20a}) yields the following conditions:
\begin{eqnarray}
&&  d_\theta {\hat {A}}(x,y)+d{\hat {C}}(x,y)+{\hat {A}}(x,y)\wedge 
       {\hat {C}}(x,y)
       +{\hat {C}}(x,y)\wedge {\hat {A}}(x,y)=0, \label{2.20a1}\\
&&  d_{\bar\theta} {\hat A}(x,y)+d{\hat{\bar C}}(x,y)
   +{\hat A}(x,y)\wedge {\hat{\bar C}}(x,y)
       +{\hat{\bar C}}(x,y)\wedge {\hat A}(x,y)=0, \label{2.20a2}\\
&&  d_\theta{\hat {{\mit\Phi}}}(x,y)+d_\chi{\hat {C}}(x,y)
       +{\hat {\mit\Phi}}(x,y)\wedge {\hat {C}}(x,y)
       +{\hat {C}}(x,y)\wedge {\hat {\mit\Phi}}(x,y)=0, \label{2.20a3}\\
&&  d_{\bar{\theta}} {\hat {\mit\Phi}}(x,y)+d_\chi{\hat{\bar {C}}}(x,y)
   +{\hat {\mit\Phi}}(x,y)\wedge {\hat{ \bar{C}}}(x,y)
       +{\hat{\bar C}}(x,y)\wedge {\hat {\mit\Phi}}(x,y)=0,
        \label{2.20a4}\\
&& d_\theta {\hat C}(x,y)+{\hat C}(x,y)\wedge{\hat C}(x,y)=0,
                                       \label{2.20a5}\\
&& d_{\bar\theta} {\hat{\bar C}}(x,y)
    +{\hat{\bar C}}(x,y)\wedge{\hat{\bar C}}(x,y)=0,  \label{2.20a6}\\
&& d_{\bar\theta} {{\hat C}}(x,y)+d_{\theta} {\hat{\bar C}}(x,y)
      +{{\hat C}}(x,y)\wedge{\hat{\bar C}}(x,y)
      +{\hat{\bar C}}(x,y)\wedge{\hat C}(x,y)=0. \label{2.20a7}
\end{eqnarray}
These determine the BRS/anti-BRS transformations of the respective fields.
Strictly speaking, $d_\theta {\hat A}(x,y)$ in Eq.(\ref{2.20a1}) 
should be written as
$d_\theta {\hat A}(x,y,\theta,{\bar\theta})|_{\theta={\bar\theta}=0}$. 
The notations used in Eqs.(\ref{2.20a2})$\sim$(\ref{2.20a7}) follows
the same convention.
Nakanishi-Lautrup fields are defined as
\begin{equation}
    d_\theta{\hat{\bar C}}(x,y)={\hat B}(x,y),\hskip 2cm 
    d_{\bar\theta}{\hat C}(x,y)={\hat{\bar B}}(x,y). \label{2.20a8}
\end{equation}
It should be noted that the nilpotencies of $d_\theta$
 and $d_{\bar\theta}$ 
are consistent with Eqs.(\ref{2.20a1})$\sim$(\ref{2.20a8}).
We would like to draw  attention 
to Eqs.(\ref{2.20a3}) and (\ref{2.20a4}), which contain 
terms of the BRST/anti-BRST of the Higgs field. 
From Eq.(\ref{2.6a2})$\sim$Eq.(\ref{2.6a4}), these two
equations can be rewritten as
\begin{eqnarray}
&&   \partial_\theta {\mit\Phi}(x,y)=\partial_y C(x,y)
+{\mit\Phi}(x,y)C(x,-y)
-C(x,y){\mit\Phi}(x,y),
                             \label{{2.20b1}}\\
&&   \partial_{\bar\theta} {\mit\Phi}(x,y)=\partial_y {\bar C}(x,y)
    +{\mit\Phi}(x,y){\bar C}(x,-y)-{\bar C}(x,y){\mit\Phi}(x,y),
                             \label{{2.20b2}}
\end{eqnarray}
which by use of the relation $H(x,y)={\mit\Phi}(x,y)+M(y)$ lead to
\begin{eqnarray}
&&   \partial_\theta H(x,y)=H(x,y)C(x,-y)-C(x,y)H(x,y),
                             \label{2.20b3}\\
&&   \partial_{\bar\theta} H(x,y)=H(x,y){\bar C}(x,-y)-
{\bar C}(x,y)H(x,y).
                             \label{2.20b4}
\end{eqnarray}
Equations (\ref{2.20b3}) and (\ref{2.20b4}) are the usual BRST/anti-BRST
of the Higgs field. \par
The BRST invariant Yang-Mills-Higgs Lagrangian$\;$ is obtained as
\begin{eqnarray}
     {\cal L}_{\rm YMH}=&&-\sum_{y=\pm}\frac1{g_y^2}
     {\rm Tr}<F(x,y), F(x,y)>\nonumber\\
             &&+\sum_{y=\pm}\frac1{g_y^2}
             i\partial_\theta\partial_{\bar\theta}
  {\rm Tr}<{\cal A}(x,y,\theta,{\bar\theta}), 
  {\cal A}(x,y,\theta,{\bar\theta})>|_{\theta={\bar\theta}=0}
   \nonumber\\
 &&+\sum_{y=\pm}\frac1{g_y^2}\frac\alpha2
           {\rm Tr}<{\hat B}(x,y,\theta,{\bar\theta}), 
   {\hat B}(x,y,\theta,{\bar\theta})>|_{\theta={\bar\theta}=0},
                \label{2.19}
\end{eqnarray}
where $g_y$ is a constant related 
to the coupling constant of the flavor gauge field, and
Tr denotes the trace over internal symmetry matrices. 
In order to express the Yang-Mills-Higgs Lagrangian$\;$, 
let us denote the explicit expressions of
the field strength $F(x,y)$.
The algebraic rules defined in Eqs.(\ref{2.2a1})$\sim$(\ref{2.2a3}), 
(\ref{2.4}) and (\ref{2.7}) yield
\begin{eqnarray}
 F(x,y) &=&{1 \over 2}F_{\mu\nu}(x,y)dx^\mu \wedge dx^\nu  \nonumber\\
           &&\hskip 1.5cm   + D_\mu {\mit\Phi}(x,y)dx^\mu \wedge \chi 
               + V(x,y)\chi \wedge \chi,
                \label{2.21}
\end{eqnarray}
where
\begin{eqnarray}
&&  F_{\mu\nu}(x,y)=\partial_\mu A_\nu (x,y) - \partial_\nu A_\mu (x,y) 
               +[A_\mu(x,y), A_\mu(x,y)], \label{2.221}\\
&&  D_\mu {\mit\Phi}(x,y)=\partial_\mu {\mit\Phi}(x,y)  
+ A_\mu(x,y)(M(y) + {\mit\Phi}(x,y))\nonumber\\
&&     \hskip 6.5cm           
 -({\mit\Phi}(x,y)+M(y))A_\mu(x,-y),\label{2.222} \\  
&&  V(x,y)= ({\mit\Phi}(x,y) + M(y))({\mit\Phi}(x,-y) + M(-y)) - Y(x,y). 
\label{2.223}
\end{eqnarray}
The function $Y(x,y)$ in Eq.(\ref{2.223})
 is an auxiliary field expressed as 
\begin{equation}
  Y(x,y)= \sum_{i}a_{i}^\dagger(x,y)M(y)M(-y)a_{i}(x,y)
 \label{2.23}
\end{equation}
which may or may not depend on ${\mit\Phi}(x,y)$ 
and/or may be a constant field.
\par
In order to obtain an explicit expression for $L_{\rm YMH}$ 
in Eq.(\ref{2.19})
we must determine the metric structure of one-forms.
\begin{eqnarray}
&& <dx^\mu, dx^\nu>=g^{\mu\nu},\quad 
g^{\mu\nu}={\rm diag}(1,-1,-1,-1),\nonumber\\
&& <\chi, \chi>=-1, \hskip 1cm 
<d\theta, d\theta>=<d{\bar\theta}, d{\bar\theta}>=1.
\label{2.25}
\end{eqnarray}
We note here that all other such combinations vanish.
From Eqs.(\ref{2.21})$\sim$(\ref{2.223}), the first term of
Eq.(\ref{2.19}), that representing 
the gauge and Higgs bosons sector, is written as
\begin{eqnarray}
{\cal L}_{\rm YMH}^{^B}
&=&-{\rm Tr}\sum_{y=\pm}{1\over 2g^2_y}
F_{\mu\nu}^{\dag}(x,y)F^{\mu\nu}(x,y)\nonumber\\
&&+{\rm Tr}\sum_{y=\pm}{1\over g_{y}^2}
    \left[D_\mu {\mit\Phi}(x,y)\right]^{\dag}D^\mu {\mit\Phi}(x,y)
      \nonumber\\
&& -{\rm Tr}\sum_{y=\pm}{1\over g_{y}^2}
        V^{\dag}(x,y)V(x,y),
\label{2.26}
\end{eqnarray}
where the third term on the right hand side of Eq.(\ref{2.26}) 
is the potential term of the Higgs particle.
The second and third terms of Eq.(\ref{2.19}) give the ghost term 
${\cal L}_{\rm GH}$ and the gauge fixing term ${\cal L}_{\rm GF}$,
respectively. 
${\cal L}_{\rm GH}$ is expressed as
\begin{eqnarray}
{\cal L}_{\rm GH}=&&-2i\sum_{y=\pm}\frac1{g_y^2}{\rm Tr}
\partial_\mu{\bar C}(x,y){\cal D}^\mu C(x,y)\nonumber\\
&&-i\sum_{y=\pm}\frac1{g_y^2}
{\rm Tr}\left[\partial_y {\bar C}(x,-y){\cal D}_yC(x,y)+
    \partial_y {\bar C}(x,y){\cal D}_yC(x,-y)\right],
            \label{2.27}
\end{eqnarray}
where 
\begin{eqnarray}
    {\cal D}^\mu C(x,y)=&&\partial^\mu C(x,y)+[A^\mu(x,y), C(x,y)], 
    \label{2.271}\\
    \partial_y{\bar C}(x,y)=&&-{\bar C}(x,y)M(y)+M(y){\bar C}(x,-y),
    \label{2.270}\\
    {\cal D}_y C(x,y)=&&\partial_y C(x,y)-C(x,y){\mit\Phi}(x,y)
    +{\mit\Phi}(x,y)C(x,-y)\nonumber\\
                     =&&-C(x,y)H(x,y)+H(x,y)C(x,-y)
               \label{2.272} 
\end{eqnarray}
and ${\cal L}_{\rm GF}$ as
\begin{eqnarray}
   {\cal L}_{\rm GF}=&&\frac\alpha2\sum_{y=\pm}\frac1{g_y^2}{\rm Tr}
   B(x,y)^2
   + 2i \sum_{y=\pm}\frac1{g_y^2}{\rm Tr}
               \partial_\mu B(x,y)A^\mu(x,y)\nonumber\\
             +&& i\sum_{y=\pm}\frac1{g_y^2}{\rm Tr}
             \left(\partial_y B(x,-y){\mit\Phi}(x,y)
             +{\mit\Phi}(x,-y)\partial_y B(x,y)\right).
               \label{2.28}
\end{eqnarray}
If we note the Hermitian conjugate conditions that 
\begin{eqnarray}
   && \left(\partial_y {\bar C}(x,y)\right)^\dagger
   =\partial_y {\bar C}(x,-y), \hskip 1cm
   \left({\cal D}_y {C}(x,y)\right)^\dagger={\cal D}_y {C}(x,-y)
    \nonumber\\
   && \left(\partial_y {B}(x,y)\right)^\dagger=-\partial_y {B}(x,-y)
                \label{2.281}
\end{eqnarray}
due to $B(x,y)^\dagger=B(x,y)$, $C(x,y)^\dagger=-C(x,y)$ and
${\bar C}(x,y)^\dagger=-{\bar C}(x,y)$, we easily find the Hermiticity of
Eqs.(\ref{2.27}) and (\ref{2.28}).
\par
There is another method to obtain the  potential 
term of the Higgs field.
Sitarz \cite{Sita} defined
the new metric $g_{\alpha\beta}$ with $\alpha$ and $\beta$ 
running over $0,1,2,3,4$ by $g^{\alpha\beta}={\rm diag}(+,-,-,-,-)$.
The fifth index here represents the discrete space $Z_2$. Then, 
$dx^\alpha=(dx^0,dx^1,dx^2,dx^3,\chi)$ is followed.
The generalized field strength $ F(x,y)$ in Eq.(\ref{2.21})
is written as $F(x,y)=F_{\alpha\beta}(x,y)dx^\alpha\wedge dx^\beta$,
 where
$F_{\alpha\beta}(x,y)$ is defined
 in Eqs.(\ref{2.221})$\sim$(\ref{2.223}). 
Then, it is
easily found that ${\rm Tr}\{g^{\alpha\beta}F_{\alpha\beta}(x,y)\}$
is gauge invariant. Thus, the term
\begin{equation}
|{\rm Tr}\{g^{\alpha\beta}F_{\alpha\beta}(x,y)\}|^2
=\{{\rm Tr}V(x,y)\}^\dagger\{{\rm Tr} V(x,y)\}    \label{2.29}
\end{equation}
can be added to Eq.(\ref{2.26}). If this term exists,
 the restriction between
coupling constants is lost, and thus,
 the Higgs mass becomes a free parameter.
\par
\section{ Application to the SU(2)$\times$U(1) gauge model}
\label{sec.3}
In this section we apply the results of the previous section
to the spontaneously broken SU(2)$\times$U(1) gauge model. 
We do not deal with the Fermion sector in this article 
because it is not related to  the main theme of this article.
Let us first assign  the fields on the discrete space $M_4\times Z_2$ 
to the fields in the SU(2)$\times$U(1) gauge model. For gauge fields
\begin{eqnarray}
&& A_\mu(x,+)=-{i \over 2}\sum_{i=1}^3\tau^i A^i_\mu(x)
           -{i \over 2}a\tau^0B_\mu(x),\nonumber\\
&& A_\mu(x,-)=-{i \over 2}bB_\mu(x),\label{3.1}
\end{eqnarray}
where $A_\mu^i(x)$ and $B_\mu(x)$ denote SU(2) and U(1) 
gauge fields, respectively. Here, $\tau^i(i=1,2,3)$ are Pauli matrices, 
and $\tau^0$ is the $2\times 2$ unit matrix. 
The condition $a-b=1$ is required because it 
corresponds to the hypercharge of the Higgs field.  The Higgs field is
assigned as
\begin{eqnarray}
&& {\mit\Phi}(x,+)={\mit\Phi}(x)=
\left(
\matrix{
\phi_+(x)\cr
\phi_0(x)\cr
}\right), \hskip 1cm 
M(+)=
\left(
\matrix{
0\cr
\mu\cr
}\right), \nonumber\\
&& {\mit\Phi}(x,-)={\mit\Phi}^{\dag}(x), \hskip 1cm 
M(-)=(0,\;\; \mu)=M^{\dag}(+),\label{3.2}
\end{eqnarray}
where $M(y)$ must be chosen to give the correct symmetry breakdown.
For ghost and anti-ghost fields, which correspond with gauge fields 
in Eq.(\ref{3.1}), we take 
\begin{eqnarray}
&& C(x,+)=-{i \over 2}\sum_{i=1}^3\tau^i C^i(x)
           -{i \over 2}a\tau^0C^0(x),\nonumber\\
&& C(x,-)=-{i \over 2}bC^0(x)\label{3.3}
\end{eqnarray}
and
\begin{eqnarray}
&& {\bar C}(x,+)=-{i \over 2}\sum_{i=1}^3\tau^i {\bar C}^i(x)
           -{i \over 2}a\tau^0 {\bar C}^0(x),\nonumber\\
&& {\bar C}(x,-)=-{i \over 2}b {\bar C}^0(x).\label{3.4}
\end{eqnarray}
Also for the Nakanishi-Lautrup field, we assign
\begin{eqnarray}
&& B(x,+)={1 \over 2}\sum_{i=1}^3\tau^i B^i(x)
           +{1 \over 2}a\tau^0B^0(x),\nonumber\\
&& B(x,-)={1 \over 2}bB^0(x)\label{3.5}
\end{eqnarray}
and
\begin{eqnarray}
&& {\bar B}(x,+)={1 \over 2}\sum_{i=1}^3\tau^i {\bar B}^i(x)
           +{1 \over 2}a\tau^0 {\bar B}^0(x),\nonumber\\
&& {\bar B}(x,-)={1 \over 2}b {\bar B}^0(x)\label{3.6},
\end{eqnarray}
because $\partial_{ \theta} {\bar C}^i=i{B}^i$ and 
$\partial_{ \theta} {\bar C}^0=i{B}^0$.
We can take the gauge transformation functions as
\begin{eqnarray}
&& g(x,+)=e^{-ia\alpha(x)}g(x),\;e^{-ia\alpha(x)}\in {\rm U(1)},\;
        g(x)\in {\rm SU(2)},\nonumber\\
&& g(x,-)=e^{-ib\alpha(x)}\in {\rm U(1)}. \label{3.7}
\end{eqnarray}
After  elimination of the auxiliary field $Y(x,+)$ and the 
rescaling of fields
\begin{equation}
 A_\mu^i(x)  \rightarrow g A_\mu^i(x),\hskip 1.5cm
 B_\mu(x) \rightarrow g'B_\mu(x),\hskip 1.5cm
 H(x) \rightarrow g_{H}H(x), \label{3.8}
\end{equation}
with $g=g_+,\;\displaystyle{g^{'}
={g_+g_-/ \sqrt{a^2g_-^2+{\displaystyle{b^2g_+^2/ 2}}}}}\,$  
and $g_{H}=g_+g_-/\sqrt{g_+^2+g_-^2}$,
we find  the standard Yang-Mills-Higgs Lagrangian$\;$ 
for the SU(2)$\times$U(1) gauge theory:
\begin{eqnarray}
{\cal L}_{\rm YMH}^{^B}=
  -{1\over 4}\left[\sum_i F_{\mu\nu}^i(x)\cdot F^{i\mu\nu}(x)
  + B_{\mu\nu}(x)\cdot B^{\mu\nu}(x)\,\right] \nonumber\\
+\left[D_\mu H(x)\right]^{\dag}\left[D^\mu H(x)\right] 
-\lambda\left[\,H^{\dag}(x)\,H(x)-\mu^2\right]^2, \label{3.9}
\end{eqnarray}
where
\begin{eqnarray}
&& F_{\mu\nu}^i (x) = \partial_\mu A_\nu^i(x)-\partial_\nu A_\mu^i(x)
       +g\,\epsilon^{ijk}A_\mu^j(x)A_\nu^j(x), \label{3.10}\\
&&  B_{\mu\nu}(x)=\partial_\mu B_\nu(x)
-\partial_\nu B_\mu(x) \label{3.10a}\\
&& D_\mu H(x)=\left[\partial_\mu-{i \over 2}
(g\,\sum_i \tau^i\cdot  A^i_\mu(x)
+g'\tau_0B_\mu(x))\right]H(x),\label{3.11}\\
&& \displaystyle{\lambda={g_+^{4}g_-^{2} 
/(g_+^2+g_-^2)^2}\,}, \hskip 1cm \mu\rightarrow g_{H}\mu. \nonumber
\end{eqnarray}
Equation(\ref{3.9}) expresses the Yang-Mills-Higgs Lagrangian$\;$ 
of the gauge theory with the symmetry
SU(2)$\times$U(1)  spontaneously broken to  SU(1)$_{\rm em}$.
If the Sitarz term in Eq.(\ref{2.29}) is added,
the coupling constant $\lambda$ is changed to $\lambda'$,  removing the
restriction between coupling constants, and therefore making  it possible
to perform the renormalization of the 
gauge theory  as usual.\par
Let us now move to the ghost 
and gauge fixing terms expressed in Eqs.(\ref{2.27}) 
and (\ref{2.28}). 
For simplicity, we hereafter omit the argument $x$ 
in the respective fields.
After applying the same rescaling of ghost and Nakanishi-Lautrup 
fields as that applied in Eq.(\ref{3.8})
\begin{eqnarray}
  &&  B^i \rightarrow gB^i, \hskip 1cm B^0 \rightarrow g'B^0, \hskip 1cm
        {\bar B}^i \rightarrow g{\bar B}^i, \hskip 1cm 
     {\bar B}^0 \rightarrow g'{\bar B}^0, \nonumber\\
  &&  C^i \rightarrow gC^i, \hskip 1cm C^0 \rightarrow g'C^0 \hskip 1cm
   {\bar C}^i \rightarrow g{\bar C}^i, \hskip 1cm 
     {\bar C}^0 \rightarrow g'{\bar C}^0, \label{3.12}
\end{eqnarray}
we obtain the gauge fixing term ${\cal L}_{\rm GF}$ in Eq.(\ref{2.27}) as
\begin{eqnarray}
  {\cal L}_{\rm GF}=&& \frac\alpha2({B_1}^2+{B_2}^2+
         {B_{Z}}^2+{B_{A}}^2) \nonumber\\
       && -B_1\left(\partial^\mu A_\mu^1-m_{W}\phi^1\right)-
        B_2\left(\partial^\mu A_\mu^2-m_{W}\phi^2\right)
        \nonumber\\
       &&-B_{Z}\left(\partial^\mu Z_\mu-m_{Z}\phi^4\right)
         -B_{A}\partial^\mu A_\mu, \label{3.13}
\end{eqnarray}
where
$A_\mu$ and $Z_\mu$ represent the photon 
and the neural weak boson, respectively,
and other fields are defined as
\begin{eqnarray}
&&    B_{A}=B_0\cos\theta_{W}+B_3\sin\theta_{W}, \hskip 1cm
      B_{Z}=-B_0\sin\theta_{W}+B_3\cos\theta_{W}, \label{3.14}\\
&&    \phi^+=\frac1{\sqrt{2}}(\phi^2+i\phi^1),\hskip 1cm 
      \phi^-=\frac1{\sqrt{2}}(\phi^2-i\phi^1), \label{3.15}\\
&&    \phi^0=\frac1{\sqrt{2}}(\phi^3-i\phi^4),\hskip 1cm 
      {\phi^0}^\ast=\frac1{\sqrt{2}}(\phi^3+i\phi^4) \label{3.16}
\end{eqnarray}
with the Weinberg angle $\theta_{W}$. The quantities
$m_{W}$ and $m_{Z}$ in Eq.(\ref{3.13}) are the charged and neutral gauge
boson masses, respectively. 
The equations of motion eliminate 
the Nakanishi-Lautrup fields from Eq.(\ref{3.13}),  yielding
\begin{eqnarray}
  {\cal L}_{\rm GF}=
       && -\frac1{2\alpha}\left(\partial^\mu A_\mu^1-
        m_{W}\phi^1\right)^2-\frac1{2\alpha}
        \left(\partial^\mu A_\mu^2-m_{W}\phi^2\right)^2
        \nonumber\\
     &&  -\frac1{2\alpha}\left(\partial^\mu Z_\mu
          -m_{Z}\phi^4\right)^2
         -\frac1{2\alpha}\left(\partial^\mu A_\mu\right)^2. \label{3.17}
\end{eqnarray}
Eq.(\ref{3.17}) enables us to obtain the gauge fixed Lagrangian
with the 't Hooft-Feynman gauge\cite{tH}. This result is 
the same  as that reported in Ref.\cite{LHN}.
\par
With the  notations used in Eq.(\ref{3.17}), 
we obtain the explicit expression of the 
ghost terms in Eq.(\ref{2.27}) as follows:
\begin{eqnarray}
       {\cal L}_{\rm GH}=&& i\partial^\mu{\bar C}^k{\cal D}_\mu C^k
                 +i\partial^\mu{\bar C}^0\partial_\mu C^0 \nonumber\\
          &&+\frac{i}{2}m_{W}
          \left\{g({\bar C}^1C^1+{\bar C}^2C^2)(\phi^3+v)+
          g({\bar C}^2C^1-{\bar C}^1C^2)\phi^4\right.\nonumber\\
         && \hskip 2.5cm
         +\left.(gC^3+g'C^0)({\bar C}^1\phi^1-{\bar C}^2\phi^2)
           \right\}\nonumber\\
        && +\frac{i}{2}m_{Z}\left\{-gC_{Z}(C^1\phi^1-C^2\phi^2)+
        \sqrt{g^2+{g'}^2}
         {\bar C}_{Z}C_{Z}(\phi^3+v)\right\}, \label{3.18}
\end{eqnarray}
where $v={\sqrt 2}\mu$, ${\cal D}_\mu C^k=\partial_\mu C^k
+\epsilon^{klm}A_\mu^lC^m$
 and 
\begin{equation}
      C_{Z}=-C_0\sin\theta_{W}+C_3\cos\theta_{W}, \hskip 1cm
      {\bar C}_{Z}=-{\bar C}_0\sin\theta_{W}+{\bar C}_3
          \cos\theta_{W}. \label{3.19}
\end{equation}
The new interaction terms between ghost and
Higgs fields appear here.
 This is a natural result because the Higgs field is
a member of the generalized connection in NCG on the discrete space
just as the gauge fields  $A_\mu, W^{\pm}_\mu$ and $Z_\mu$.

\section{ Concluding remarks}
The BRST invariant Lagrangian of the spontaneously broken gauge theory is 
presented in our scheme by introducing the Grassmann numbers $\theta$ and
${\bar \theta}$ as the arguments in super space in addition to 
$x_\mu$ in $M_4$ and $y$ in $Z_2$. The horizontality condition prescribes 
the BRS transformations of the respective fields,
 including the Higgs field.
By use of the generalized gauge field
 ${\cal A}(x,y,\theta,{\bar \theta})$
and the Nakanishi-Lautrup field ${\bar B}(x,y,\theta,{\bar \theta})$
as in Eq.(\ref{2.19}), 
the gauge fixing and ghost terms appear in the Lagrangian, 
yielding the t'Hooft-Feynman gauge as the gauge fixing condition and
extra interactions between ghost and Higgs fields. \par
If we include in the Yang-Mills-Higgs Lagrangian$\;$ 
the quartic term of the Higgs field due to Sitarz
\cite{Sita} in Eq.(\ref{2.29}), 
it attains a form which is
 free from any restriction on coupling constants, 
making it possible to quantize the BRST invariant Lagrangian 
in the same way as 
the ordinary Lagrangian. Generally speaking, 
the NCG approach to clarify the Higgs mechanism 
 provides  the  estimations of the Weinberg angle and 
the mass relation between 
the Higgs and gauge bosons or top quark.  
These estimations  may be in a sense inconsistent
 with the quantization of 
the theory. However, some of these relations\cite{Soga},\cite{OFS} 
are too attractive to be discarded.
Alvarez, Gracia-Bond${\acute {\rm i}}$a and 
Mart${\acute {\rm i}}$n\cite{AGM} in fact calculated 
such NCG constraints using the renormalization group analysis
 and obtained
interesting relations, including $m_{top}=2m_{W}$ 
and $m_{H}=3.14m_{W}$,
where the NCG constraints are assumed
 to hold at some renormalization point.
In order to perform such analyses, a renormalizable Lagrangian including
the gauge fixing and ghost terms must exist. In this sense, the work of 
Lee, Hwang and Ne'eman\cite{LHN} 
is very important for such renormalization
analysis. The present work also shows how  we can 
obtain the renormalizable Lagrangian. In a previous work\cite{OFS}, 
we found the relations $\sin^2\theta_{_{W}}=3/8$
and $m_{_{H}}=\sqrt{2}m_{_{W}}$, assumed to be satisfied 
at the energy of the GUT scale. 
In addition, we  obtained the mass relation 
$\displaystyle{m_H=\frac4{\sqrt{3}}m_W\sin\theta_W}$
 with the assumption that
the coupling constant of the Higgs quartic term is not  affected
by the Sitarz term. It might be expected in the NCG scheme that 
the quadratic divergent term of the Higgs self-energy
 disappears because of  gauge invariance,
if the Higgs field would
 be the genuine gauge field on the discrete space.
If this were true, the mass relation 
$\displaystyle{m_H=\frac4{\sqrt{3}}m_W\sin\theta_W}$ might hold
without any correction in the same way as $m_W=m_Z\cos\theta_W$.
In any case, it is very interesting 
to investigate the renormalization analysis 
of these results according to the BRST invariant Lagrangian presented
in this article. This will be done in a future work.

\section*{ Acknowledgements}
The author would like to
express his sincere thanks to
Professors J.~Iizuka,
 H.~Kase, K. Morita and M.~Tanaka 
for useful suggestions and
invaluable discussions on  noncommutative geometry.
I am also grateful to Dr. G. Paquette for reading the manuscript and
correcting  inappropriate expressions.

\end{document}